\documentclass[11pt]{article}
\newfont{\larom}{cmbx9 scaled\magstep3}
\newfont{\bsan}{cmssbx10}
\begin{document}
\begin{center}
  {\larom Limited Frequency Range Observations of Cosmological
	  Point Sources}

  \vspace{5mm}
  {\Large Marcelo B. Ribeiro \\}
  \vspace{5mm}
  {\normalsize \it Vatican Observatory Group, Steward Observatory,
    University of Arizona \\ and \\
    Physics Institute, University of Brazil - UFRJ, Rio de Janeiro \\} 

  \vspace{8mm}
  {\bf ABSTRACT}
\end{center}
  \begin{quotation}
    \small
    This paper advances a general proposal for testing non-standard
    cosmological models by means of observational relations of
    cosmological point sources in some specific waveband, and their
    use in the context of the data provided by the galaxy redshift
    surveys, but for any cosmological metric. By starting from the
    general theory for observations in relativistic cosmology
    the equations for colour, K-correction, and number counts of
    cosmological point sources are discussed in the context of curved
    spacetimes. The number counts equation is also written in terms
    of the selection and luminosity functions, which provides a
    relativistic generalization of its Euclidean version. Since
    these observables were not derived in the framework of any
    specific cosmology, they are valid for {\it any} cosmological
    model. The hypotheses used in such derivation are reviewed,
    together with some difficulties for the practical use of those
    observables.
    
    \vspace{3mm}

  \end{quotation}
\subsection*{Introduction}

The standard Friedmann models are generally considered as the best
approximation for the observed large scale distribution of galaxies,
since the results predicted by these models are usually quite good
approximations of observations.$^1$ However, although
no observational evidence was so far found to severely contradict this
widespread belief, the question remains of whether or not other
cosmological models could also provide theoretical predictions in
line with observations. This is obviously an important aspect in the
general acceptance of the standard Friedmannian models as good
approximations of the observed Universe, inasmuch as we can only have
a direct response to the question of how good the Friedmann models
really are, if we are able to test the data against the predictions
of other non-standard cosmological models.

Nevertheless, cosmography is presently dominated by observational
relations derived only within the Friedmannian context,$^{1-4}$ and
obviously those relations do not allow comparisons between standard
and non-standard cosmologies. Therefore, in practice we presently
have a situation where the observational test of non-standard models
is quite difficult due to the absence of detailed and
observationally-based relations derived for that purpose.

There are exceptions, however, and the basis of a general theory for
observations of cosmological sources was presented by George Ellis,$^5$
although later, in a series of papers,$^{6-8}$ the theory was further
developed, with the presentation of detailed calculations of
observational relations from where cosmological effects can be 
identified and separated from the brightness profile evolution of the
sources.

Although such study was a step forward in the possibility of direct
observational test of non-standard cosmological models, this detailed
theory$^{6-8}$ equally demands detailed observations of the sources, a
task usually not feasible when dealing with large scale redshift surveys,
where the total number of observed objects varies from hundreds to
thousands of galaxies. Actually, often it is not even desirable to
obtain such detailed observations since what is often being 
sought are data for doing statistics on the distribution of galaxies.

The approach of this work differs from those quoted above because 
in here cosmological sources are considered point sources, and
therefore observables like flux and colour are integrated over
the whole object. This is a reasonable approximation for 
objects included in these surveys, since they are usually so faint
that very detailed observations of their structure are still 
difficult with the presently available techniques. Therefore, by
treating galaxies as point sources we can, at least in principle,
apply the methods presented in this paper to the large and deep
galaxy surveys presently available.

The observational relations discussed here were derived with the
aim of comparing with this redshift surveys of galaxies. As a
consequence, the theory used here offers the possibility of comparing the
predictions of different cosmological models with the need of much
less real data than demanded by the detailed theory mentioned
above.$^{6-8}$ Besides, this simpler view of the problem creates the
option of a first order test of cosmological models against
observations without the need of detailed data, which in turn would
demand a more complex and demanding analysis. However, in order to
be able to obtain observational relations capable of being compared
with observations, to a certain extent we need to depart from the basic
approach$^5$ and discuss in detail some specific observations in
cosmology within some specific bandwidth, since this is the way
astronomers deal with their data.

This paper is the first of a series in a programme for investigating
whether or not other, non-standard, cosmological models could also
explain the data obtained from the large-scale redshift surveys of
galaxies. Here I shall review the basic theory for observational
relations in limited frequency bandwidth, and the quantities which
are mostly used by observers. In doing so I will put together some
basic results which will form the common ground from where the
general approach of this proposed research programme should start. I
will also extend some aspects of this theory, particularly the
number-counts expression, and indicate
where the connection among these observational quantities, real
astronomical observations, and the spacetime geometry takes place.
In short, such a connection appears when the observables are written
in terms of the redshift and the cosmological distances, since both
can only be explicitly written when a spacetime metric is assumed.
Even when all observables are written solely in terms of the redshift,
this connection will appear in the functional form between the
observational quantities and the redshift, as this functional
relationship is dependable on the chosen spacetime geometry.

\subsection*{Basic Definitions and Equations}

Let us call $F$ the {\it bolometric flux} as measured by the observer.
This is the rate at which radiation crosses unit area per unit time
in all frequencies. Then $F_{\scriptscriptstyle G}$ will be the
{\it bolometric galaxy flux} measured across an unit sphere located
in a locally Euclidean space at rest with a galaxy or a cosmological
source.$^5$

The distance definitions used here are three: {\it i}) the {\it observer
area distance} $r_0$ is the area distance of a source as measured by the
observer; {\it ii}) the {\it galaxy area distance}
$r_{\scriptscriptstyle G}$ is defined as the area distance to the
observer as measured from the distant galactic source. This quantity is
unobservable, by definition; {\it iii}) the {\it luminosity distance}
$d_\ell$ is the distance measured by the observer as if the space were
flat and non-expanding, that is, as if the space were stationary and
Euclidean. The observer area distance $r_0$ is also called {\it angular
diameter distance},$^2$ and {\it corrected luminosity distance}.$^{11}$
The galaxy area distance $r_{\scriptscriptstyle G}$ is also named
{\it effective distance},$^{12}$ {\it angular size distance},$^1$
{\it transverse comoving distance},$^{13}$ and {\it proper motion
distance}.$^{14}$ These three definitions of distance are related to
each other by {\it Etherington's reciprocity theorem},$^{5,15,16}$ 
  \begin{equation}
     d_\ell = r_0 \; {(1+z)}^2 = r_{\scriptscriptstyle G} \; (1+z),
     \label{distancias}
  \end{equation}
where $z$ is the {\it redshift} of the source. All these distances tend
to the same Euclidean value as $z \rightarrow 0$, but greatly differ at
large redshift.$^{9,10}$ 

Although the equation above appears in standard texts of observational
cosmology, with very few exceptions$^{5,16}$ they all fail to
acknowledge the generality of the theorem, and give due credit to
Etherington's 1933 discovery. The reciprocity relation was proven for
general null geodesics, without specifying {\it any} metric, and,
therefore, it is {\it not at all} restricted to standard cosmologies.

Let us now call $L$ the {\it bolometric source luminosity}, that is, the
total rate of radiating energy emitted by the source and measured
through an unity sphere located in a locally Euclidean spacetime
near the source. Then $\nu$ will be the {\it observed frequency}
of the radiation, and $\nu_{\scriptscriptstyle G}$ the {\it emitted
frequency}, that is, the frequency of the same radiation $\nu$
received by the observer, but at rest-frame of the emitting galaxy.

The {\it source spectrum function} ${\cal J} (\nu_{\scriptscriptstyle G})$
gives the proportion of radiation emitted by the source at a certain
frequency $\nu_{\scriptscriptstyle G}$ as measured at the rest frame
of the source. This quantity is a property of the source, giving the
percentage of emitted radiation, and obeying the normalization
condition, $ \int_0^\infty {\cal J} (\nu_{\scriptscriptstyle G}) d
\nu_{\scriptscriptstyle G} = 1$. Then $L_{\nu_{\scriptscriptstyle G}} =
L \ {\cal J}(\nu_{\scriptscriptstyle G})$ is the {\it specific source
luminosity}, giving the rate at which radiation is emitted by the
source at the frequency $\nu_{\scriptscriptstyle G}$ at its locally
Euclidean rest frame. To summarize, we have that,
\begin{equation}
   L = \int_{\mbox{\rm \tiny 2-sphere}} F_{\scriptscriptstyle G} dA
     = 4 \pi F_{\scriptscriptstyle G}
     = \int_0^\infty L_{\nu_{\scriptscriptstyle G}} d
       \nu_{\scriptscriptstyle G}
     = \int_0^\infty L {\cal J} (\nu_{\scriptscriptstyle G}) d
       \nu_{\scriptscriptstyle G}.
   \label{fluxo}
\end{equation}
The redshift $z$ is defined by,
  \begin{equation} 1+ z = \frac{\lambda_{\mbox{\scriptsize observed}}}
     {\lambda_{\mbox{\scriptsize emitted}}}
     = \frac{\nu_{ \scriptscriptstyle G}}{\nu},
     \label{redshift}
   \end{equation}
and from the expressions above it follows that
   \begin{equation}
      d \nu = \frac{ d \nu_{\scriptscriptstyle G}}{(1+z)},
      \label{2}
   \end{equation}
and
   \begin{equation}
      F = \frac{F_{\scriptscriptstyle G}}{{(r_0)}^2 {(1+z)}^4}
	= \frac{F_{\scriptscriptstyle G}}{{(r_{\scriptscriptstyle
	  G})}^2 {(1+z)}^2}
        = \frac{F_{\scriptscriptstyle G}}{{(d_\ell)}^2}.
      \label{fluxos}
   \end{equation}

The connection of the model with the spacetime geometry
appears in the expressions for the redshift and the different
definitions of distance. That can be seen if we remember that in the
general geometric case the redshift is given by,$^5$ 
\begin{equation}
 1+z = \frac{{\left( u^a k_a \right) }_{\mbox{\scriptsize source}}}
 {{\left( u^a k_a \right) }_{\mbox{\scriptsize observer}}},
 \label{z_em_ellis}
\end{equation}
where $u^a$ is the observer's four-velocity, and $k^a$ is the tangent
vector of the null geodesic connecting source and observer, that is, the
past light cone. This expression allows us to calculate
$z$ for any given spacetime geometry. If we assume that source and
observer are comoving, then $u^a=\delta^a_0$ implies that
$u^bk_b=k^bg_{0b}$, and the redshift may be rewritten as,
\begin{equation}
   1+z= \frac{{[ g_{0b} ( dx^b/dy ) ] }_{\mbox{\scriptsize source}}}
             {{[ g_{0b} ( dx^b/dy ) ] }_{\mbox{\scriptsize observer}}}.
   \label{n2}
\end{equation}
Here $y$ is the affine parameter along the null geodesics connecting
source and observer, and $g_{ab}$ is the metric tensor. Inasmuch as
$dx^b/dy$ and $g_{ab}$ can only be determined when a spacetime geometry
is defined by some line element $dS^2$, the function $g_{0b}(dx^b/dy)$
and, ultimately, the redshift as well are directly dependable on the
geometry of the model. Although $z$ is an astronomically
observable quantity, its specific internal relationship with other
internal cosmological quantities of the model will be set by the metric
tensor.

The observer area distance $r_0$ is defined by,$^{5,16}$
\begin{equation}
  (r_0)^2 = \frac{d A_0}{d \Omega_0},
  \label{area_distance}
\end{equation}
where $d A_0$ is the cross-sectional area of a bundle of null geodesics
measured {\it at the source's rest frame}, and diverging from the observer
at some point, and $d \Omega_0$ is the solid angle subtended by this
bundle. This quantity can in principle be measured if we had {\it
intrinsic} astrophysically-determined dimensions of the source, but it
can also be obtained from the assumed spacetime geometry, especially in
spherically symmetric metrics, from where it can be easily calculated.
For the Einstein-de Sitter cosmology, detailed calculations for many
observables can be found elsewhere.$^{10,17}$

\subsection*{Frequency Bandwidth Observational Relations}

\subsubsection*{\it Flux and Magnitude}

The flux  within some specific wavelength range can be obtained
if we consider equations (\ref{fluxo}), (\ref{2}) and (\ref{fluxos}).
Then we have,$^5$
\begin{equation}
  F= \frac{\int_0^\infty L {\cal J} (\nu_{\scriptscriptstyle G})
      d \nu_{\scriptscriptstyle G}}
     {4 \pi {(r_0)}^2 {(1+z)}^4} = \frac{L}{4 \pi} \frac{\int_0^\infty
     {\cal J} \left[ \nu (1+z) \right] (1+z) d \nu}{{(r_0)}^2 {(1+z)}^4} =
     \frac{L}{4 \pi}\frac{\int_0^\infty {\cal J} \left[ \nu (1+z) \right]
     d \nu}{{(r_0)}^2 {(1+z)}^3}.
     \label{4}
\end{equation}
Therefore, the {\it specific flux} $F_{\nu}$ measured in the frequency
range $\nu$, $\nu + d \nu$ by the observer, may be written as
\begin{equation}
   F_{\nu} d \nu = \frac{L}{4 \pi} \frac{{\cal J} \left[ \nu (1+z)
   \right] d \nu} {{(r_0)}^2 {(1+z)}^3}.
  \label{5}
\end{equation} 
The apparent magnitude in a specific observed frequency bandwidth is, 
\begin{equation}
  m_{\scriptscriptstyle W} = - 2.5 \log \int_0^\infty F_\nu W(\nu) d \nu +
  \mbox{\rm constant},
  \label{mag2}
\end{equation}
where $W( \nu)$ is the function which defines the spectral interval
of the observed flux (the standard UBV system, for instance). This is
a sensitivity function of the atmosphere, telescope and detecting device.
Thus, from equations (\ref{5}) and (\ref{mag2}) the apparent magnitude in
a specified spectral interval $W$ yields,
\begin{equation}
  m_{\scriptscriptstyle W} = -2.5 \log \left\{ \frac{L}{4 \pi}
  \frac{1}{{(r_0)}^2 {(1+z)}^3} \int_0^\infty W(\nu) {\cal J}
  \left[ \nu (1+z) \right] d \nu \right\} + \mbox{\rm constant}.
  \label{mag-delta}
\end{equation}

Since cosmological sources do evolve, the intrinsic luminosity $L$
changes according to the evolutionary stage of the source, and
therefore, $L$ is actually a function of the redshift: $L=L(z)$. Hence,
in order to use equation (\ref{mag-delta}) to obtain the apparent
magnitude evolution of the source, some theory for luminosity evolution
is also necessary. For galaxies, $L(z)$ is usually derived taking into
consideration the theory of stellar evolution, from where some simple
equations for luminosity evolution can be drawn.$^{1,18}$ Note that
equation (\ref{mag-delta}) also indicates that the source spectrum
function ${\cal J}$ might evolve and change its functional form at
different evolutionary stages of the source. In addition, as ${\cal J}
\left[ \nu (1+z) \right] $ is a property of the source at a specific
redshift, this function must be known in order to calculate the apparent
magnitude, unless the K-correction approach is used (see below).
For magnitude limited catalogues, the luminosity distance and the observer
area distance have both an upper cutoff, which is a function of the
apparent magnitude, the frequency bandwidth used in the observations
and the luminosity of the sources.

\subsubsection*{\it K-Correction}

The relations above demand the knowledge of both the source spectrum
and the redshift. However, when the source spectrum is not known, it
is necessary to introduce a correction term in order to obtain the
bolometric flux from observations. This correction is known as the
K-correction, and it is a different way for allowing the effect of
the source spectrum.

In deriving the K-correction,$^{3,4,19,20}$ I start by calculating the
difference in magnitude produced by the bolometric flux $F$ and the
flux $F_{\scriptscriptstyle W}$ measured by the observer, but at the
bandwidth $W(\nu)$ in any redshift $z$. Since,
\begin{equation}
  F = \int_0^\infty F_\nu d \nu, \ \ \ \ F_{\scriptscriptstyle W} =
  \int_0^\infty F_\nu W(\nu) d \nu,
  \label{fluxes}
\end{equation}
the difference in magnitude $\Delta m (z)$ will be given by
\begin{equation}
  \log \frac{F(z)}{F_{\scriptscriptstyle W}(z)} = 0.4 \Delta m (z).
  \label{delta-m}
\end{equation}
The rate between the observed flux $F_{\scriptscriptstyle W}(z)$ at
a given redshift and at $z=0$ defines the K-correction. Then,
considering equation (\ref{delta-m}), we have that
\begin{equation}
 \frac{F_{\scriptscriptstyle W}(z)}{F_{\scriptscriptstyle W}(0)}
 = \frac{F(z)}{F(0)} 10^{-0.4 K_{\scriptscriptstyle W}},
 \label{18}
\end{equation}
where we have defined
\begin{equation}
   K_{\scriptscriptstyle W} \equiv \Delta m (z) - \Delta m (0).
   \label{19}
\end{equation}
Then it follows that
\begin{equation}
   K_{\scriptscriptstyle W} = m_{\scriptscriptstyle W} -
   m_{\mbox{\scriptsize bol}} - \Delta m (0),
   \label{20}
\end{equation}
which means that once we know the K-term and the observed magnitude
$m_{\scriptscriptstyle W}$, the bolometric magnitude is know within a
constant $\Delta m (0)$. If we now substitute equation (\ref{5})
into equation (\ref{18}), and assume $L(z)=L(0)$, it is easy to show
 that 
\begin{equation}
   K_{\scriptscriptstyle W} (z) = 2.5 \log \left\{ \frac{\int_0^\infty
   W(\nu) {\cal J} (\nu) d \nu }{ \int_0^\infty W(\nu)
   {\cal J} ( \nu_{\scriptscriptstyle G})d \nu_{\scriptscriptstyle G}}
  \right\}.
   \label{21}
\end{equation}
Remembering that by equation (\ref{2}) we know that we can have
the source spectrum transformed from the rest frame of the source
to the rest-frame of the observer by a factor of $(1+z)$, that is,
${\cal J} \left[ \nu (1+z) \right] d \nu = \left[ {\cal J}
(\nu_{\scriptscriptstyle G}) d \nu_{\scriptscriptstyle G} \right]
/ (1+z) $, then we may also write equation (\ref{21}) as
\begin{equation}
  K_{\scriptscriptstyle W} (z) = - 2.5 \log (1+z) + 2.5 \log \left\{
  \frac{\int_0^\infty W(\nu) {\cal J} (\nu) d \nu }{ \int_0^\infty
  W(\nu) {\cal J} \left[ \nu (1+z) \right] d \nu } \right\}.
 \label{22}
\end{equation}

Note that the equations above allow us to write theoretical
K-correction expressions for any given spacetime geometry, provided
that the line element $d S^2$ is known beforehand. As a final remark,
it is obvious that if the source spectrum is already known, all
relevant observational relations can be calculated without the need
of the K-correction.

\subsubsection*{\it Colour}

With the expressions above we can obtain the theoretical equation
for the colour of the sources for any given spacetime. Let us consider
two bandwidths $W$ and $W'$. From equation (\ref{mag-delta}) we can find
the difference in apparent magnitude for these two frequency bands in
order to obtain an equation for the colour of the source in a
specific redshift. Let us call this quantity
$C_{{\scriptscriptstyle W}{\scriptscriptstyle W'}}$. Thus,
\begin{equation}
  C_{{\scriptscriptstyle W}{\scriptscriptstyle W'}}
  (z) \equiv
  m_{\scriptscriptstyle W} - m_{\scriptscriptstyle W'}
  = 2.5 \log \left\{ \frac{\int_0^\infty W' (\nu) {\cal J}
  \left[ \nu (1+z) \right] d \nu }{ \int_0^\infty W(\nu)
  {\cal J} \left[ \nu (1+z) \right] d \nu } \right\}.
 \label{colour1}
\end{equation}

Considering that cosmological sources do evolve, they should emit 
different luminosities in different redshifts due to the different
evolutionary stages of the stellar contents of the sources, and this
is reflected in the equation above by the source spectrum function
which may be different for different redshifts. Note, however, that
in the equation above the source is assumed to have the same bolometric
luminosity in a specific redshift and, therefore, we can only use
equation (\ref{colour1}) to compare observation of objects of the same
class and at similar evolutionary stages in certain $z$, since $L=L(z)$.
This often means galaxies of the same morphological type. In other
words, equation (\ref{colour1}) assumes that a homogeneous
populations of cosmological sources do exist, and hence, the
evolution and structure of the members of such a group will be similar.

Equation (\ref{colour1}) also gives us a method for assessing the
possible evolution of the source spectrum. For instance, by calculating
$(B-V)$ and $(V-R)$ colours for E galaxies with modern determinations of
the K-correction, it has been reported$^4$ that no colour
evolution was found to at least $z=0.4$. However, for $z \ge 0.3$
it was found that rich clusters of galaxies tend to be bluer (the
Butcher-Oemler effect) than at lower redshifts.$^{1,21}$ Therefore,
if we start from a certain metric, we can calculate the theoretical
redshift range where colour evolution would be most important for the
assumed geometry of the cosmological model. Then, assessing evolution
could be done by means of multicolour observations. As the luminosity
and area distances must be the same in all wavelengths for each given
source, if the luminosity-redshift plot is not the same in two colours,
this shows that these two colours have different evolution functions.
Applications of this idea for searching inhomogeneities, by means of
the Lema\^{\i}tre-Tolman-Bondi cosmology, can be found in the
literature.$^{22}$ 

Another point worth mentioning, from equation (\ref{colour1}) we
see that colour is directly related to the intrinsic
characteristics of the source, its evolutionary stage, as given by the
redshift and the assumptions concerning the real form of the source
spectrum function at a certain $z$. However, this reasoning is valid for
point sources whose colours are integrated and, therefore, we are not
considering here structures, like galactic disks and halos,
which in principle may emit differently and then will produce
different colours. If we remember that cosmological sources are usually
far enough to make the identification and observation of source
structures an observational problem for large scale galaxy surveys,
this hypothesis seems reasonable  at least as a first approximation.

Finally, it is clear that in order to obtain a relationship
between apparent magnitude and redshift we need some knowledge about the
dependence of the intrinsic bolometric luminosity $L$ and the source
spectrum function ${\cal J}$ with the redshift. It seems that such a
knowledge must come from astrophysically independent theories
about the intrinsic behaviour and evolution of the sources, and not
from the assumed cosmological model.

\subsubsection*{\it Number Counts}

In any cosmological model if we consider a small affine parameter 
displacement $dy$ at some point P on a bundle of past null geodesics
subtending a solid angle $d \Omega_0$, and if $n$ is the number density
of radiating sources per unit proper volume at P, then the {\it number
of sources} in this section of the bundle is,$^5$ 
\begin{equation}
  dN = {(r_0)}^2 d \Omega_0 { \left[ n ( -k^a u_a ) \right] }_P \ dy,
  \label{number1}
\end{equation}
where $k^a$ is the propagation vector of the radiation flux. Equation
(\ref{number1}) assumes the counting of {\it all} sources at P
with number density $n$. Consequently, if we want to consider the
more realistic situation that only a fraction of galaxies in the
proper volume $dV = {(r_0)}^2 d \Omega_0 d l = {(r_0)}^2 d
\Omega_0 ( - k^a u_a ) dy$ is actually detected and included in the
observed number count, we have to write $dN$ in terms of a {\it
selection function} $\psi$ which represents this detected fraction
of galaxies. Then equation (\ref{number1}) becomes$^{23}$ 
\begin{equation}
 dN_0 = {\psi} dN = {\psi} \ { \left[ n dV \right] }_P = {(r_0)}^2
	\ {\psi} \ d \Omega_0 { \left[ n ( - k^a u_a ) \right] }_P \ dy,
 \label{number2}
\end{equation}
where $dN_0$ is the {\it fractional number of sources actually observed}
in the unit proper volume $dV$ with a total of $dN$ sources.

In principle ${\psi}$ can be estimated from a knowledge of the
galactic spectrum, the observer area distance, the redshift, and
the detection limit of the sample as given by the limiting flux in
a certain frequency bandwidth. The other quantities in equation
(\ref{number2}) come from the assumed cosmological model itself.

In order to determine $\psi$ we need to remember that in any
spacetime geometry the observed flux in bandwidth $W$ is given by
equations (\ref{5}) and (\ref{fluxes}),
\begin{equation}
  F_{\scriptscriptstyle W} = \frac{L(z)}{4 \pi {(r_0)}^2 {(1+z)}^3}
    \int_0^\infty W(\nu) {\cal J} \left[ \nu (1+z) \right] d \nu.
  \label{26}
\end{equation}
Then, if a galaxy at a distance $r_0$ is to be seen at flux 
$F_{\scriptscriptstyle W}$, its luminosity $L(z)$ must be bigger
than $ \{ 4 \pi {(r_0)}^2 {(1+z)}^3 F_{\scriptscriptstyle W} \}
/ \{ \int_0^\infty W(\nu) {\cal J} \left[ \nu (1+z) \right] d \nu
\}$. Therefore, the probability that a galaxy at a distance $r_0$
and with redshift $z$ is included in a catalog with maximum
flux $F_{\scriptscriptstyle W}$ is,
\begin{equation}
  {\cal P} \propto \psi (\ell) =
  \int_{\ell}^\infty \phi (w) dw,
  \label{27}
\end{equation}
where this integral's lower limit is 
\begin{equation}
  \ell = \frac{1}{L_\ast} \frac{4 \pi {(r_0)}^2 {(1+z)}^3
	F_{\scriptscriptstyle W} (z) }{\int_0^\infty W(\nu)
	{\cal J} \left[ \nu (1+z) \right] d \nu },
  \label{28}
\end{equation}
$L_\ast$ is a parameter, and $\phi (w)$ is the {\it luminosity
function}.$^1$ In Schechter$^{24}$ model $L_\ast$ is a {\it
characteristic luminosity} at which the luminosity function exhibits
a rapid change in its slope. Now, if we assume spherical symmetry,
then equation (\ref{number2}) becomes,
\begin{equation}
  dN_0 = 4 \pi {(r_0)}^2 \ \psi (\ell) \ { \left[ n
         ( - k^a u_a ) \right] }_P \ dy.
  \label{29}
\end{equation}
Thus, the number of galaxies observed up to an affine parameter
$y$ at a point P down the light cone, may be written as
\begin{equation}
  N_0 = 4 \pi \int_0^y {(r_0)}^2 \ \psi (\ell)
	\ { \left[ n ( - k^a u_a ) \right] }_P \ d {\bar{y}},
  \label{30}
\end{equation}
which generalizes Peebles' Euclidean equation (7.40)$^{1}$ into
a relativistic setting.

Equation (\ref{30}) is deceptively simple. It is in fact a highly
non-linear and difficult-to-compute function, as all quantities
entering the integrand are functions of the past null cone affine
 parameter $y$. Therefore, in principle, they must be
explicitly calculated before they can be entered into equation
(\ref{30}). In some cases one may avoid this explicit determination
and use instead the radial coordinate,$^{10,17,25-28}$ a method which
turns out to be easier than finding these expressions in terms of $y$.
Then, once $N_0(y)$ is obtained, it becomes possible to relate it to
other observables, since they are all function of the past null cone
affine parameter. For example, if one can derive an analytic expression
for the redshift in a given spacetime, say $z=z(y)$, and if this
expression can be analytically inverted, then we can write $N_0$ as a
function of $z$.

It is important to mention that the local number density $n$
is given in units of proper density and, therefore, in order to take a
proper account of the curved spacetime geometry, one must relate $n$
to the local density as given by the right hand side of Einstein's
field equations. If, for simplicity, we suppose that all sources
are galaxies with a similar rest-mass $M_g$, then $n= {\rho}/{M_g}$.

The discussion above shows that the theoretical
determination of $N_0$ depends critically on the spacetime geometry
and the luminosity function $\phi$. For the latter, in the
Schechter$^{24}$ model it has the form, $ \phi (w) = \phi_\ast
w^\alpha e^{-w}$, where $\phi_\ast$ and $\alpha$ are constant
parameters. One must not forget that this luminosity function shape
was originally determined from {\it local} measurements,$^{24}$ and
it is still under assessment the possible change of shape and parameters
of the luminosity function in terms of evolution,$^{29-32}$
that is, as we go down the light cone.

As a final remark, one must note that gravitational lensing
magnification can also affect the counting of point sources,
because weak sources with low flux might appear brighter due to
lensing magnification. Such an effect will not be discussed here,
since its full treatment demands more detailed information about
the sources themselves, such as considering them as extended ones,
and is considered to be most important for QSO's.$^{16}$ 

\subsection*{Conclusion}

In this paper I have advanced a general proposal for testing
non-standard cosmological models by means of observational relations
of cosmological point sources in some specific waveband, and their
use in the context of data provided by the galaxy redshift surveys.
I have also shown how the relativistic number-counting equation
can be expressed in terms of the selection and luminosity functions,
generalizing thus the Euclidean number counts expression into a relativistic
setting. The equations for colour, and K-correction were also
presented. All expressions obtained here are valid for any cosmological
metric, since no specific geometry was assumed in such a derivation.
Although these observables can be specialized for a given spacetime
geometric, some quantities must come from astrophysical considerations,
namely the intrinsic luminosity $L(z)$, the source spectrum function
${\cal J}(\nu)$, and the luminosity function $\phi (w)$. These cannot be
obtained only from geometrical considerations, which means that the
determination of the spacetime structure of universe is a task
intrinsically linked to astrophysical considerations and results.
Further developments and applications of the general approach discussed
here are the subject of a forthcoming paper.$^{33}$ 

 \begin{flushleft}
 {\large \bf Acknowledgments}
 \end{flushleft}
I am grateful to W.\ R.\ Stoeger for reading the original manuscript
and for helpful comments. Financial support from Brazil's CAPES
Foundation is also acknowledged.
\begin{flushleft}
{\large \bf References}
\end{flushleft}
\begin{description}
 \item [{\rm (1)}] P.J.E.\ Peebles, {\it Principles of Physical Cosmology}
       (Princeton University Press), 1993.
 \item [{\rm (2)}] S.\ Weinberg, {\it Gravitation and Cosmology} (Wiley,
       New York), 1972.
 \item [{\rm (3)}] A.\ Sandage, {\it ARA\&A}, {\bf 26}, 561, 1988.
 \item [{\rm (4)}] A.\ Sandage, in B.\ Binggeli \& R.\ Buser (eds.), {\it
       The Deep Universe (Saas-Fee Advanced Course 23)} (Springer, Berlin),
       1995, p.\ 1.
 \item [{\rm (5)}] G.F.R.\ Ellis, in R.K.\ Sachs (ed.), {\it General
       Relativity and Cosmology (Proc.\ Int.\ School Phys.\ ``Enrico
       Fermi'')} (Academic Press, New York), 1971, p.\ 104.
 \item [{\rm (6)}] G.F.R.\ Ellis \& J.J.\ Perry, {\it MNRAS}, {\bf 187},
       357, 1979.
 \item [{\rm (7)}] G.F.R.\ Ellis, J.J.\ Perry \& A.W.\ Sievers, {\it AJ},
       {\bf 89}, 1124, 1984.
 \item [{\rm (8)}] A.W.\ Sievers, J.J.\ Perry \& G.F.R.\ Ellis, {\it MNRAS},
       {\bf 212}, 197, 1985.
 \item [{\rm (9)}] G.C.\ McVittie, {\it QJRAS}, {\bf 15}, 246, 1974.
 \item [{\rm (10)}] M.B.\ Ribeiro, {\it Gen.\ Rel.\ Grav.}, {\bf 33}, 1699,
       2001 (astro-ph/0104181).  
 \item [{\rm (11)}] J.\ Kristian \& R.K.\ Sachs, {\it ApJ}, {\bf 143},
       379, 1966.
 \item [{\rm (12)}] M.S.\ Longair, in B.\ Binggeli \& R.\ Buser (eds.),
       {\it The Deep Universe (Saas-Fee Advanced Course 23)} (Springer,
       Berlin), 1995, p.\ 317.
 \item [{\rm (13)}] D.\ W.\ Hogg, {\it preprint}, 2000 (astro-ph/9905116).
 \item [{\rm (14)}] D.\ Scott, J.\ Silk, E.W.\ Kolb \& M.S.\ Turner,
         in A.N.\ Cox (ed.), {\it Allen's Astrophysical Quantities},
	 4th edition (Springer, Berlin), 2000, p.\ 643.
 \item [{\rm (15)}] I.M.H.\ Etherington, {\it Phil.\ Mag.}, {\bf 15}, 761,
       1933; reprinted, {\it Gen.\ Rel.\ Grav.}, in press, 2002.
 \item [{\rm (16)}]P.\ Schneider, J.\ Ehlers \& E.E.\ Falco, {\it
         Gravitational Lenses} (Springer, Berlin), 1992.
 \item [{\rm (17)}] M.B.\ Ribeiro, {\it ApJ}, {\bf 441}, 477, 1995
       (astro-ph/9910145). 
 \item [{\rm (18)}] J.\ Binney \& S.\ Tremaine, {\it Galactic Dynamics}
       (Princeton University Press), 1987.
 \item [{\rm (19)}] M.L.\ Humason, N.U.\ Mayall \& A.R.\ Sandage, {\it ApJ},
       {\bf 61}, 97, 1956.
 \item [{\rm (20)}] J.B.\ Oke \& A.\ Sandage, {\it ApJ}, {\bf 154}, 21, 1968.
 \item [{\rm (21)}] R.G.\ Kron, in B.\ Binggeli \& R.\ Buser (eds.), {\it
       The Deep Universe (Saas-Fee Advanced Course 23)} (Springer, Berlin),
       1995, p.\ 233.
 \item [{\rm (22)}] C.\ Hellaby, {\it A\&A}, {\bf 372}, 357, 2001
       (astro-ph/0010641).
 \item [{\rm (23)}] G.F.R.\ Ellis, S.D.\ Nel, W.R.\ Stoeger \& A.P.\
       Whitman, 1985, {\it Phys.\ Rep.}, {\bf 124}, 315, 1985.
 \item [{\rm (24)}] P.\ Schechter, {\it ApJ}, {\bf 203}, 297, 1976.
 \item [{\rm (25)}] M.B.\ Ribeiro, {\it ApJ}, {\bf 388}, 1, 1992. 
 \item [{\rm (26)}] M.B.\ Ribeiro, {\it ApJ}, {\bf 395}, 29, 1992.
 \item [{\rm (27)}] M.B.\ Ribeiro, {\it ApJ}, {\bf 415}, 469, 1993. 
 \item [{\rm (28)}] M.B.\ Ribeiro, in D.\ Hobbil, A.\ Burd \& A.\ Coley
       (eds.), {\it Deterministic Chaos in General Relativity} (Plenum,
       New York), 1994, p.\ 269.
 \item [{\rm (29)}] C.J.\ Lonsdale \& A.\ Chokshi, {\it AJ}, {\bf 105},
       1333, 1993. 
 \item [{\rm (30)}] C.\ Gronwall \& D.C.\ Koo, {\it ApJ}, {\bf 440}, L1,
       1995.
 \item [{\rm (31)}] R.S.\ Ellis {\it et al.}, {\it MNRAS}, {\bf 280}, 235,
       1996.
 \item [{\rm (32)}] N.\ Cross {\it et al.}, {\it MNRAS}, in press, 2002
       (astro-ph/0012165). 
 \item [{\rm (33)}] M.B.\ Ribeiro \& W.R.\ Stoeger, in preparation, 2002.
\end{description}
\end{document}